\documentclass[12pt]{article}

\usepackage{graphicx}
\usepackage{dcolumn}
\usepackage{bm}
\usepackage{subfigure}

\begin{document}

\thispagestyle{empty}

\begin{flushright}
UK/11-05
\end{flushright}

\begin{center}

{\Large\bf Charge-dependent Azimuthal Correlations in Relativistic\\
\bigskip
Heavy-ion Collisions and Electromagnetic Effects } \\

\vspace{1.0cm}
{\bf K.F. Liu} \\

\bigskip
{\it
Dept.\ of Physics and Astronomy, University of Kentucky, Lexington, KY 40506}
\end{center}

\begin{abstract}
     We propose a scenario where the pattern of the recently observed charge-dependent azimuthal correlations can be
understood qualitatively. This is based on the cluster picture and the assumption that the charged hadrons that 
flow outward from the surface of the overlapping region of the colliding nuclei move primarily parallel to the reaction plane.
We also point out the there is a strong electric field induced by the transient magnetic field during the parton
production in the initial phase of the relativistic heavy-ion collision and discuss its possible relevance to the scenario.
\end{abstract}

\vspace*{2cm}

\section{Introduction}  \label{intro}

    Relativistic heavy-ion collision provides a fertile ground for a subtle interplay between electrodynamics and QCD, since
the magnetic field is very strong and short-lived so that a strong electric field is induced according to 
Faraday's law. Recently, the charge-dependent azimuthal correlations were detected by STAR at
Relativistic Heavy-Ion Collider~\cite{STAR09a} in an attempt to detect the chiral magnetic effect (CME) due to a 
possible local strong CP violation~\cite{kha98,kmw08}. Since all the direct parity-odd observables would vanish due 
to the fact that the induced current in CME can either align or anti-align with the angular momentum, the STAR experiment
measured the correlator $\langle \cos(\phi_{\alpha} + \phi_{\beta} - 2 \Psi_{RP})\rangle$, where $\phi_{\alpha}$ and
$\phi_{\beta}$ are the azimuthal angles of the charged particles and $\Psi_{RP}$ is that of the reaction plane.
This quantity is not parity-violating~\cite{STAR09a,kr10}, but can be sensitive to parity-violating effect.
It is shown by STAR~\cite{STAR09a} that in the three-particle correlator, the explicit determination of the
reaction plane is not needed~\cite{STAR04}. Assuming that particle $c$ in the three-particle correlation is
correlated to particles $\alpha$ and $\beta$ only in common correlation to the reaction plane, one has
$\langle  \cos(\phi_{\alpha} + \phi_{\beta} - 2 \phi_c)\rangle =
   \langle \cos(\phi_{\alpha} + \phi_{\beta} - 2 \Psi_{RP})\rangle v_{2,c},$
where $v_{2,c}$ is the elliptic flow of the particle $c$.

\begin{figure}[ht]
  \centering
 \includegraphics[width=18cm,height=10cm]{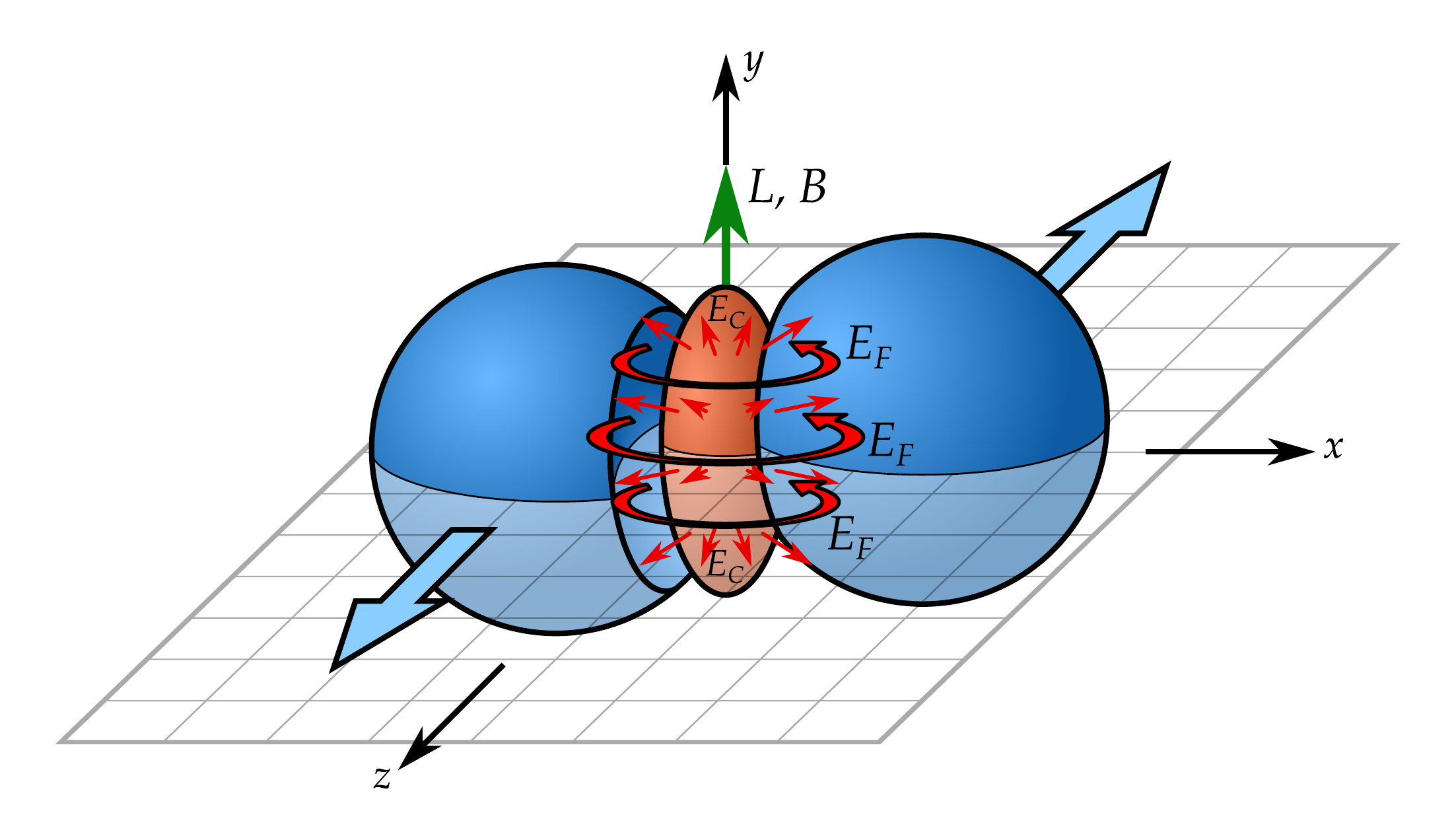}
\caption{(color online) The reaction is taken to be along the $x-z$ plane so that the angular momentum $\vec{L}$ and
 the magnetic field $\vec{B}$ in the center of the colliding nuclei are along the $y$ axis. The electric field $\vec{E_F}$
(in red) induced by the transient magnetic field due to the Faraday law is circling around the $y$ axis. The red arrows
pointing outward indicate the electric field $\vec{E}_C$ due to the electric charges which is weak in the overlapping region because 
of the cancellation of the Coulomb potentials from the charges in the two colliding nuclei.}  \label{EMRHIC}
\end{figure}

   Based on this assumption, the experimental three-particle azimuthal correlation and the separately measured
correlation $\langle  \cos(\phi_{\alpha} - \phi_{\beta})\rangle$ are combined to obtain the correlations of products
of the azimuthal angles $\langle \sin(\phi_{\alpha})\sin(\phi_{\beta})\rangle$ and
$\langle \cos(\phi_{\alpha})\cos(\phi_{\beta})\rangle$ for the like-sign charged pairs (+,+), (-,-) and
unlike-sign charged pairs (+,-), (-,+)~\cite{bkl10}. The frame is chosen in such a way that the reaction plane is 
in the $x-z$ plane
and the $y$ axis is perpendicular to the reaction plane. We shall adopt the same frame which is illustrated in
Fig.~\ref{EMRHIC}. In this case, both the angular momentum $\vec{L}$ and the central magnetic field $\vec{B}$ are along the
the positive $y$ axis.
The salient features revealed from this analysis~\cite{bkl10} for these charge-dependent azimuthal correlations are 
the following:

\begin{itemize}
\item
      For the like-sign (LS) charged pairs, it is found
\begin{eqnarray}
& \langle \sin(\phi_{\alpha})\sin(\phi_{\beta})\rangle_{LS} \simeq 0, \label{LSS}\\
&  \langle \cos(\phi_{\alpha})\cos(\phi_{\beta})\rangle_{LS} < 0. \label{LCC}
\end{eqnarray}

\item
     For the unlike-sign (US) charged pairs, on the other hand, one finds the following relation
\begin{equation} \label{USC}
\langle \sin(\phi_{\alpha})\sin(\phi_{\beta})\rangle_{US} \simeq  \langle \cos(\phi_{\alpha})\cos(\phi_{\beta})\rangle_{US} > 0.
\end{equation}
\end{itemize}

     Furthermore, the smallness of $\langle \sin(\phi_{\alpha})\sin(\phi_{\beta})\rangle_{LS} $ in
Eq.~(\ref{LSS}) holds for the range of centrality from 65\% (peripheral collision) down to $\sim 0\%$ (cental collision);
while $\langle \cos(\phi_{\alpha})\cos(\phi_{\beta})\rangle_{LS}$ is increasingly more negative from central
to peripheral collision. On the other hand, $\langle \sin(\phi_{\alpha})\sin(\phi_{\beta})\rangle_{US}$ and
$\langle \cos(\phi_{\alpha})\cos(\phi_{\beta})\rangle_{US}$ for the unlike-sign cases are increasingly more positive
from central to peripheral collision. They are almost identical from 0\% to $\sim 60\%$ centrality and
are about a factor of two larger in magnitude as compared to $\langle \cos(\phi_{\alpha})\cos(\phi_{\beta})\rangle_{LS}$
(Eq.~(\ref{LCC})) in this centrality range.

     The smallness of $\langle \sin(\phi_{\alpha})\sin(\phi_{\beta})\rangle_{LS} $ is taken to imply that
the like-sign charged pairs are mostly in plane, since $\sin(\phi)$  is non-negative for all the azimuthal angles~\cite{bkl10}. 
The fact that
$\langle \cos(\phi_{\alpha})\cos(\phi_{\beta})\rangle_{LS}$ is negative is taken to suggest~\cite{bkl10} that the
the like-sign pairs are not only in plane but more back-to-back than in the same $x$-direction (N.B. the azimuthal angle is
measured from the reaction plane $x-z$ in reference to positive $x$ (See Fig.~\ref{outward}). This seems to be in contradiction with
CME which predicts that the like-sign pairs are out of plane. However, the signals of CME could be smaller than the
background events due to the strong electric and magnetic fields and the omnipresent strong interaction. In this case, it
is essential to understand the background so that one can hope to pull out the signal of CME. In the case of unlike-sign
pairs, the positive $\langle \cos(\phi_{\alpha})\cos(\phi_{\beta})\rangle_{US}$ suggests that they are more in the
same $x$ direction, i.e. with the relative azimuthal angle $|\phi_{\alpha} - \phi_{\beta}| < 90^{\circ}$. However, it is hard
to comprehend why $\langle \sin(\phi_{\alpha})\sin(\phi_{\beta})\rangle_{US}$ and
$\langle \cos(\phi_{\alpha})\cos(\phi_{\beta})\rangle_{US}$ should be the same~\cite{bkl10}. Since there is an elliptic
flow, one would expect $\langle \cos(\phi_{\alpha})\cos(\phi_{\beta})\rangle_{US}$ which favors the in-plane flow to be
larger than $\langle \sin(\phi_{\alpha})\sin(\phi_{\beta})\rangle_{US}$ which favors out-of-plane flow. Moreover, it is not
clear as to why the sizes of  $\langle \sin(\phi_{\alpha})\sin(\phi_{\beta})\rangle_{US}$ and
$\langle \cos(\phi_{\alpha})\cos(\phi_{\beta})\rangle_{US}$ are a factor of two larger than that of
$\langle \cos(\phi_{\alpha})\cos(\phi_{\beta})\rangle_{LS}$, with the latter being negative.

     We shall present a scenario which may allow us to understand the azimuthal correlations of charged hadrons in a
qualitative manner. The observation of clustering and its implications will be discussed in Sec.~\ref{cluster}. The
time dependence of the magnetic field and the induced electric field in the initial phase for the nuclear collision 
and their impacts on the initial parton momentum direction are presented in Sec.~\ref{EM}. Partly motivated by
the large electric field present in the early phase of the collision, we postulate in Sec.~\ref{cdac} that the azimuthal correlations mainly 
originate from the clusters on the surface of the overlapping region and proceed to calculate the charge-dependent
azimuthal correlations. We end with a conclusion in Sec.~\ref{conclusion}.

\section{Clustering}   \label{cluster}

     We first address the observation that the azimuthal correlations in Eqs.~(\ref{LCC}) and (\ref{USC}) increase
from the more central collisions to peripheral ones, i.e. as \% of centrality increases. This can be largely understood as due to
clustering. In measuring the rapidity and azimuthal correlations of two charged particles produced in the pp collision in
the central rapidity region, it was observed~\cite{EFT75} that if one multiplies $n-1$, where n is the number of charged 
particles, to the azimuthal correlations, they turn out to be independent of $n$ within errors. This is an indication of 
short-range
correlation which can be described by independent emission of low-multiplicity clusters. The $n-1$ dependence comes from the
fact that the correlation scales with $n$ which is proportional to the number of clusters, while the total charged particles
scales with $n(n-1)$. The same analysis based on the cluster model was carried out for Cu\, +\, Cu and Au\, +\, Au collisions at 
center of mass energy of 200 GeV~\cite{PHO-NN}. On top of the $n-1$ scaling observed in the pp case, there is an additional
suppression of effective size (multiplicity) of the cluster toward more central collision, and more so with the away-side
($90^{\circ}< \Delta\phi < 180^{\circ}$) pairs than with the near-side ($0^{\circ}< \Delta\phi < 90^{\circ}$)
pairs. It is suggested that in more central collisions, the correlated decays from the clusters, in space-time, inside the 
dense overlapping region get lost and only the correlations from the clusters on the surface survive. Thus, it is the
combined effects of  $n-1$ scaling and the survival of correlations from the clusters on the surface which are responsible for 
the observed decrease of the azimuthal correlations toward more central collisions, as the centrality is measured by the inelastic 
cross section or the total multiplicity of particle production. 

  Following this natural explanation of the centrality dependence, we shall explore the possibility of understanding
the azimuthal correlations in Eqs.~(\ref{LSS})-(\ref{USC}) in terms of the cluster description where the correlated pairs
emerge from within those clusters which are mostly on the surface of the overlapping region. 
The effects of cluster particle correlations on two- and three-particle azimuthal correlations and their parameters
have been studied from the STAR experimental input~\cite{wangfq10}. We shall present a scenario based on the cluster
model to explain the azimuthal correlations themselves. 

\section{Electric and Magnetic fields}  \label{EM}

  Next, we consider the time dependence of the electric and magnetic field landscape in the initial phase of the nuclear collision 
at relativistic speed as well as the parton production during this period. It is well-known that there is a strong magnetic field 
from the colliding relativistic heavy ions. For the relevant overlapping
region, it is mostly directed in the direction of the angular momentum which is perpendicular to the reaction plane as
depicted in Fig.~\ref{EMRHIC}. This has been studied in the Hadron-String-Dynamics (HSD) model~\cite{vtc11}. For example, 
in $A_u + A_u$ collision at $\sqrt{S_{NN}} = 200$ GeV, and impact parameter b = 10 fm, the magnetic field $e B_y$ at the center rises from zero at 
$t ~ - 0.1$ fm/c (N.B. $t$ = 0 is refereed to the time when the two nuclei are maximally overlapped.) to a maximum of 
$\sim 4.5\, m_{\pi}^2$ at $t = 0.05$ fm/c and quickly decays to $\sim 0.7\, m_{\pi}^2$ at t = 0.15 fm/c~\cite{vtc11}. On the
other hand, due to the cancellation from the two nuclei, the electric field due to the Coulomb potential of the charges 
is weak in the overlapping region which we depict as $E_C$ in Fig.~\ref{EMRHIC}. For example, in $A_u + A_u$ 
collision at $\sqrt{S_{NN}} = 200$ GeV, b = 10 fm, the peak electric fields in the x-direction $eE_x \sim 0.5\, m_{\pi}^2$ 
at $x = 1 -3 $ fm and t = 0.05 - 0.1 fm/c. This is an order of magnitude smaller than the peak magnetic field in the center 
of the overlapping region at t = 0.05 fm/c. However, we should point out that there is another transient electric field which has not 
been taken into account. Since the strong magnetic field decreases very quickly~\cite{kmw08,vtc11}, i.e. $eB_y$ decreases 
maximally from $\sim 4.5\, m_{\pi}^2$ at t = 0.05 fm/c to $\sim 0.7\, m_{\pi}^2$ at t = 0.15 fm/c~\cite{vtc11}. According to Faraday's law, 
this rapidly changing magnetic field leads to a transient electric field $E_F$ which is circling the $y$-axis in the $x-z$ 
plane (depicted in Fig.~\ref{EMRHIC}) with a magnitude 
\begin{equation}
|eE_F| = \frac{1}{2 \pi r} \frac{\Delta \phi_B}{\Delta t} = 19\, r(fm)\, m_{\pi}^2,
\end{equation}
where $r$ in fermi is the distance from the $y$-axis. For $b = 10$ fm and the radius of the $Au$ \mbox{$R = 1.2 A^{1/3} = 7$ fm},
the surface of the overlapping region around the reaction plane is at $r = 3$ fm. At this distance from the center,
$|eE_F|$ in the $x-z$ plane is $\sim 57\, m_{\pi}^2$ which
is an order of magnitude larger than the maximum magnetic field and is even several times larger than the central NN force 
near the core at 0.7 fm of the Argonne V18 potential~\cite{wss95}. With such a strong, albeit short-lived, electric field $E_F$, the 
charged partons produced at this early stage of the collision ($\sim 0.05$ fm/c) can acquire an impact with
an initial momentum along the reaction plane as large as \mbox{$|\Delta p_{\|}| = |eE_F| \Delta t = 560$ MeV/c.} This is comparable to the
average $p_{\perp} \sim 1$ GeV/c~\cite{ew94}. In order to have a more realistic picture, one needs to fold in the
timeline of the parton production during this initial phase. Using the Heavy Ion Jet INteraction Generator (HIJING) 
model to illustrate the space-time evolution of the parton production~\cite{ew94}, it is found that the partons are mostly produced
in the initial phase of the collision before 0.5 fm/c and the `semi-hard' partons are produced between $t = -0.15$ fm/c and
0.15 fm/c. The total parton number produced as a function of time for the central collision of $Au$ on $Au$ at
$\sqrt{S} = 200$ GeV is plotted in Fig. 3 in Ref.~\cite{ew94}. Before the nuclei overlap at $t=0$, most of the partons are produced 
via initial state bremsstrahlung. Since these radiations are almost collinear, these partons move along the beam direction and have
large rapidities ($y$ is narrowly peaked at 2 at $t =0$). These partons with large longitudinal momentum will leave the interaction
region after semi-hard scattering and will not rescatter with the beam in leading twist. Since the STAR experiment of
the azimuthal correlation are measured in the mid-pseudorapidity region ($|\eta| < 1.0$) and $0.15 \leq p_t \leq 2$ GeV/c, these
partons produced before $t=0$ hardly will contribute to the experimental measurements. Between $t = 0$ and $t = 0.2$ fm/c where the
magnetic field $eB_y$ varies from 3 $m_{\pi}^2$ at $t =0$ to a maximum of 4.5 $m_{\pi}^2$ at $t = 0.05$ fm/c and down to
$0.2\, m_{\pi}^2$ at $t=0.2$ fm/c, the partons are produced via semi-hard scattering and the rate of production is the largest. The 
semi-hard scattering produces partons uniformly in the rapidity range $y < 2$ and fill up the mid and central rapidity region. These partons
are copiously produced in the duration where the magnetic field varies the most and will be imparted large momentum from the
induced electric field parallel to the reaction plane. Those produced at $t \sim 0$ and afterward will see a net decrease of $eB_y$ and,
thus, will experience a net electric force. Those on the surface of the overlap region, the maximum momentum transfer is 560 MeV/c 
as estimated above. After the semi-hard scattering, the final state radiation will also raise the mid and central rapidity of the 
produced partons, yet the magnetic field does not decrease much ($eB_y$ changes from 0.2 $m_{\pi}^2$ at $t = 0.2$ fm/c down to
near zero at $t = 0.3$ fm/c.) Thus, these partons will not be affected much by the electric force. But, the
total partons produced after semi-hard scattering (i.e. after $t = 0.2$ fm/c) is small ($\sim 20$\% of the total)~\cite{ew94}.
To sum up this part, the majority of the partons produced with rapidity $y < 1$ are produced via semi-hard scattering and it
coincides with the large decrease of the magnetic field (from  3 $m_{\pi}^2$ at $t =0$ to $0.2\, m_{\pi}^2$ at $t=0.2$ fm/c) where
the induced electric field $E_F$ gives a large momentum transfer in the direction parallel to the reaction plane. 
 Moreover, the magnetic field, which is perpendicular to the reaction plane, will also exert 
a force on the moving charged partons along the reaction plane. In other words, both the
electric field $E_F$ and the magnetic field exert strong forces on the charged partons (quarks)
along the reaction plane according to the equation $\vec{F} = q \vec{E}_F + q \vec{v} \times \vec{B}$. 

\section{Charge-dependent Azimuthal Correlations}   \label{cdac}

\begin{figure}[ht]
  \centering
  \subfigure[] 
     {\label{outward}
     {\includegraphics[width=8cm,height=5cm]{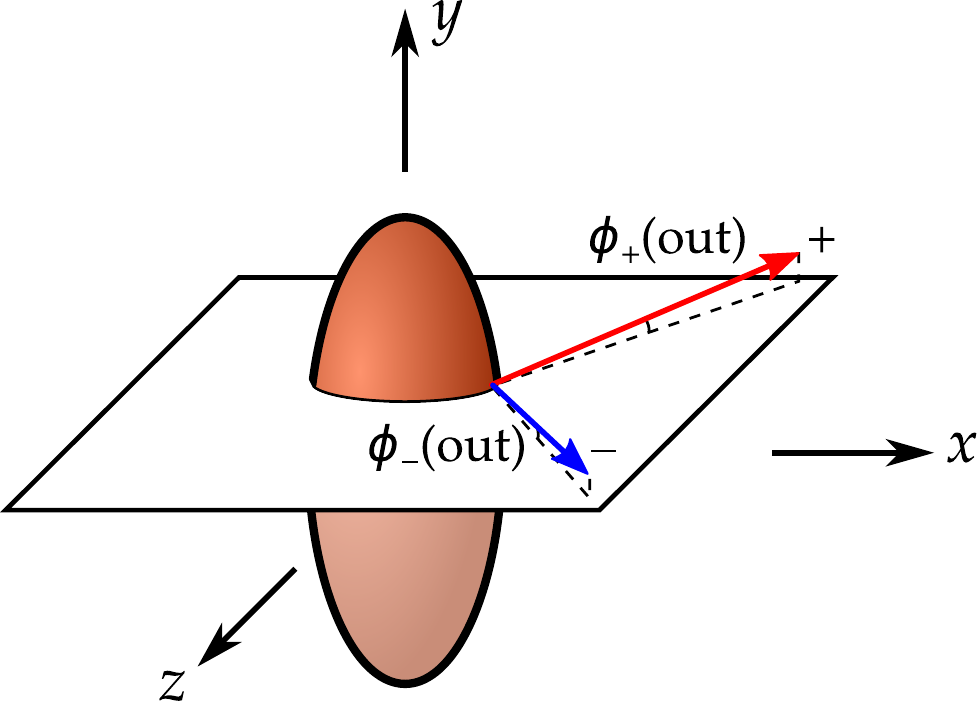}\ \ \ \ }}
  \hspace{0.6cm}
  \subfigure []
     {\label{2-pairs}
     {\includegraphics[width=6.5cm,height=4.5cm]{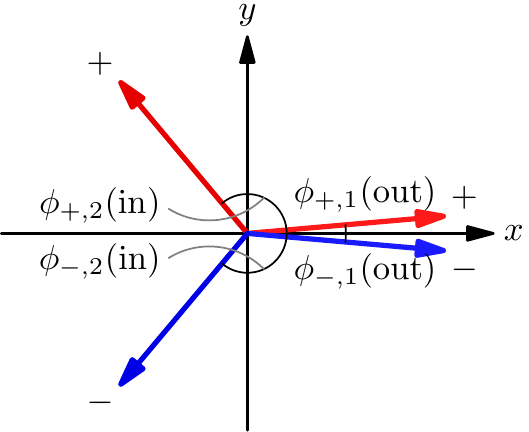}\ \ \ \ }}
  \caption{(color online) (a) An illustration of the outward going charged hadron pairs which are mainly parallel to the
       reaction plane so that the azimuthal angles $\phi_+(out)$ and $\phi_-(out)$ are small. (b) Two charged pairs 
       originating from one cluster with one pair going outward from the overlapping region and the other going 
       inward through the overlapping nuclear medium.}
\end{figure}

     We don't know how much the induced electric field will affect the subsequent hadronization and rescattering of the emergent 
charged hadrons. It will require dynamical modeling by incorporating the electromagnetic effect in this early stage as the initial 
condition for a more detailed description of the ensuing transport. In fact, there are calculations of the charge-dependent azimuthal 
correlations 
based on the Hadron String Dynamics (HSD) transport~\cite{vtc11} and A Multi-Phase-Transport (AMPT) model~\cite{mz11} which found that 
the azimuthal correlations are much smaller than found experimentally. However, these calculations have not included the effect of 
the induced electric field from the Faraday law. It would be interesting to see if the induced electric field could have a large effect in 
these calculations. Partly motivated by the presence of this large transient electric field which imparts momentum transfer to
the partons along the reaction plane initially as discussed above, we shall make an
ansatz that the azimuthal correlations mainly originate from the clusters on the surface of the overlapping region as discussed 
earlier and, furthermore, the positive- and negative-charged hadron 
pairs emerging outward from these surface  clusters, as depicted in Fig.~\ref{outward}, will move 
mainly along the reaction plane. This parallel motion could be enhanced by the subsequent Coulomb electric field when 
the non-overlapping part of the charged nuclei fly apart~\cite{vtc11}. This is
illustrated in Fig.~\ref{outward} where the positive- and negative-charged hadron pairs emerging from 
a surface cluster in the positive $x$-hemisphere of the overlapping region with the positively charged particle momentum in
the $(x,y,z)= (+,\pm,-)$ direction and negatively charged particle momentum in the
$(x,y,z)= (+,\pm,+)$ direction. But both have small azimuthal angles relative to the reaction plane with 
$\sin \phi \sim \epsilon$. On the other hand, the charged hadron pairs from a surface cluster which were originally produced from
the partons heading toward the inner region will interact strongly with the denser medium~\cite{kmw08} and emerge in the negative 
$x$-direction with large azimuthal angles as illustrated in Fig.~\ref{2-pairs}.

     It is learned that the effective size of the cluster $M_{cluster}$ is $\sim 3$ in the Cu\, +\, Cu and Au\, +\, Au collisions at 
center of mass energy of 200 GeV~\cite{PHO-NN}. The average number of pairs is
\mbox{$\langle N_{pair} (N_{pair} -1)/2 \rangle = M_{cluster}^2/2$} so that the total number of pairs is~\cite{wangfq10}
\begin{equation}
N_{LS} + N_{US} = N_{cluster} M_{cluster}^2/2,
\end{equation}
from the Poisson distribution.  Here, $N_{LS}$ is the number of like-sign (LS) pairs and  $N_{US}$ is the number of 
unlike-sign (US) pairs. With $M_{cluster} \sim 3$, the average sum of $N_{LS}$ and $N_{US}$ is $\sim 4.5$ for each cluster.
We shall consider the case from 4 up to 8 such pairs of $N_{LS} + N_{US}$ with the same number of positive- and negative-charged 
hadrons emitted from each cluster.

Now let's consider the case of 2 positive and negative pairs from a cluster according to the ansatz -- 
one pair in the positive $x$-direction with small 
azimuthal angles and the other one in the negative $x$-direction with large azimuthal angles in Fig.~\ref{2-pairs}. 
For the like-sign (LS) pairs,
$\langle \sin \phi_{\pm,1}(\rm{out}) \sin \phi_{\pm,2}(\rm{in})\rangle_{LS} \sim O(\epsilon)$ and
$\langle\cos \phi_{\pm,1}(\rm{out}) \cos \phi_{\pm,2}(\rm{in})\rangle_{LS} < 0$ in this case. These are in agreement with 
Eqs.~(\ref{LSS}) and (\ref{LCC}). For the unlike-sign (US) pair,
there are three terms for $((+,1)(\rm{out}), (-,1)(\rm{out})), ((+,2)(\rm{in}), (-,2)(\rm{in}))$ and 
$((+,1)(\rm{out}), (-,2)(\rm{in}))$ and $((-,1)(\rm{out}), (+,2)(\rm{in}))$ pairs
\begin{eqnarray}
\langle \sin \phi_{+,1}(\rm{out}) \sin \phi_{-,1}(\rm{out})\rangle_{US} &\sim& O(\epsilon^2); \nonumber \\
\langle \sin \phi_{\pm,1}(\rm{out}) \sin \phi_{\pm,2}(in)\rangle_{US} &\sim& O(\epsilon); \nonumber \\
\langle \sin \phi_{+,2}(\rm{in}) \sin \phi_{-,2}(\rm{in})\rangle_{US} & = S_{US} > 0,
\end{eqnarray}
where
\begin{equation}  \label{S_US}
S_{US} = (\frac{2}{\pi})^2 \int_{\pi/2}^{\pi} d\phi_{\alpha}d\phi_{\beta}
\,\, \sin \phi_{\alpha} \sin \phi_{\beta}\, f_{US}(\phi_{\alpha}, \phi_{\beta}),
\end{equation}
with $f_{US}(\phi_{\alpha}, \phi_{\beta})$ being the distribution of $\phi_{\alpha}$ and $\phi_{\beta}$ for
$\pi/2 \leq \phi_{\alpha}, \phi_{\beta} \leq \pi$. 
Since $sine$ is semi-positive definite, $S_{US}$ is positive which is in agreement with
Eq.~(\ref{USC}) in sign.

\begin{figure}[ht]
  \centering
  \subfigure[] 
     {\label{3a}
     {\includegraphics[width=6.5cm,height=4.5cm]{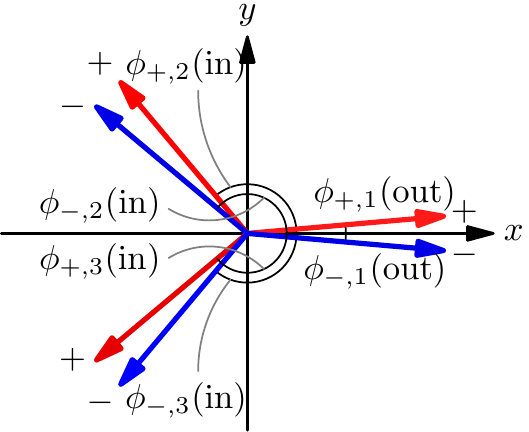}\ \ \ \ }}
  \hspace{0.6cm}
  \subfigure []
     {\label{3b}
     {\includegraphics[width=6.5cm,height=4.5cm]{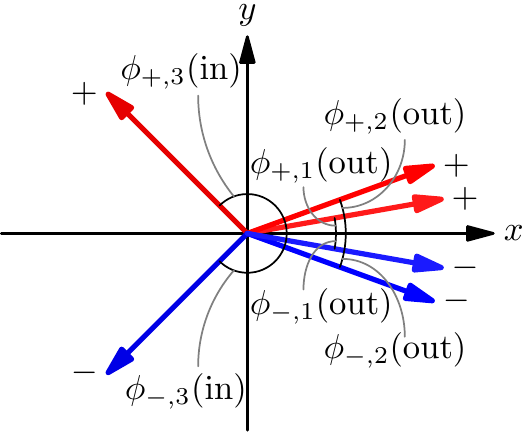}\ \ \ \ }}
  \caption{(color online) Three pairs of positive- and negative-charged particles with one pair going outward and 
   two pairs inward in (a) and two pairs going outward and one pair inward in (b).}
  \label{}
\end{figure}

Similarly,
\begin{eqnarray}
\langle \cos \phi_{+,1}(\rm{out}) \cos \phi_{-,1}(\rm{out})\rangle_{US} &\sim& 1; \nonumber \\
\langle \cos \phi_{\pm,1}(\rm{out}) \cos \phi_{\mp,2}(in)\rangle_{US}&=& 2 c\, < 0; \nonumber \\
\langle \cos \phi_{+,2}(\rm{in}) \cos \phi_{-,2}(\rm{in})\rangle_{US} & =& C_{US}\, > 0,
\end{eqnarray}
where
\begin{eqnarray}
 c &=& \frac{2}{\pi} \int_{\pi/2}^{\pi} d\phi_{\alpha}\,\, \cos \phi_{\alpha}(\rm{in})  g(\phi_{\alpha}); \\
C_{US} &=& (\frac{2}{\pi})^2 \int_{\pi/2}^{\pi} d\phi_{\alpha}d\phi_{\beta}
\,\, \cos \phi_{\alpha}(\rm{in}) \cos \phi_{\beta}(\rm{in})\, f_{US}(\phi_{\alpha}, \phi_{\beta}).
\label{C_US}
\end{eqnarray}
Both angles $\phi_{\alpha}(\rm{in})$ and $\phi_{\beta}(\rm{in})$ are between $\pi/2$ and
$\pi$; therefore, $C_{US} > 0$.
To the extent that $ (1 + 2c + C_{US}) > 0$, it would be in agreement with Eq.~(\ref{USC}) in sign. 
Taking a crude estimate that the inward particles emerging in the negative $x$-hemisphere have a uniform 
distribution and $\phi_{\alpha}(\rm{in})$ and $\phi_{\beta}(\rm{in})$ are uncorrelated, then $c = - 2/\pi$ and  
$C_{US} = 4/(\pi)^2$. In this case,  $(1 + 2c + C_{US}) = (1 - 2/\pi)^2 > 0$.

We can extend this scenario to more general cases with equal numbers of positively/negatively 
charged particles ($M_+ = M_- = M$) in the
positive $x$-hemisphere and similarly with $N_+ = N_- = N$ charged particles in the negative 
$x$-hemisphere, all originate from one cluster. For the LS case
\begin{equation}
\langle \sin \phi_{\alpha} \sin \phi_{\beta}\rangle_{LS} \propto  M(M -1)\epsilon^2
+ 2 MN \epsilon\, s + N(N-1) S_{LS},
\end{equation}
where $s = \int_{\pi/2}^{\pi} d\alpha \sin \phi_{\alpha}(\rm{in}) g(\alpha)$ gives the
contribution from the cross term with one particle in the $+x$ hemisphere with
$\sin \phi (out) \sim \epsilon$ and the other one in the $-x$ hemisphere with a distribution
of $g(\alpha)$. $S_{LS}$ is the same as $S_{US}$ in Eq.~(\ref{S_US}) except the
distribution function $f_{US}$ is replaced with $f_{LS}$.
Similarly, 
\begin{equation}
\langle \cos \phi_{\alpha} \cos \phi_{\beta}\rangle_{LS} \propto  M(M -1)
+ 2 MN\,c + N(N-1) C_{LS},
\end{equation}
where $C_{LS}$ is the same as $C_{US}$ in Eq.~(\ref{C_US}) except the
distribution function $f_{US}$ is replaced with $f_{LS}$.

On the other hand, for the unlike sign case 
\begin{eqnarray}
\langle \sin \phi_{\alpha} \sin \phi_{\beta}\rangle_{US} &\propto&  M^2 \epsilon^2
+ 2 MN \epsilon\,s + N^2 S_{US}, \\
\langle \cos \phi_{\alpha} \cos \phi_{\beta}\rangle_{US} &\propto&  M^2
+ 2 MN\,c + N^2 C_{US}.
\end{eqnarray}

     We give a table of these results for the cases with a total number of
charged particles $K$ ($K = 2 (M+N)$) being from 4 to 8 in each cluster in Table~\ref{list}.

\begin{table}[h]
\caption{Contributions to charge-dependent azimuthal correlations for 4 to 8 charged particles from
2 to 4 pairs of positive- and negative-charged hadrons.
 \label{list}}
\begin{center}
\begin{tabular}{|c|cc|c|c|c|c|}
  \hline
   Total Numbers & M & N & $\langle \sin \phi_{\alpha} \sin \phi_{\beta}\rangle_{LS}$ & $\langle 
\cos \phi_{\alpha} \cos \phi_{\beta}\rangle_{LS}$ & $\langle \sin \phi_{\alpha} \sin \phi_{\beta}\rangle_{US}$ &
  $\langle \cos \phi_{\alpha} \cos \phi_{\beta}\rangle_{US}$ \\
  {\footnotesize(Charged Particles)} & & & & & &\\
\hline \hline
   4  & 1 & 1 & $O(\epsilon)$ & 2 c  & $S_{US}$  & 1 + 2c + $C_{US}$ \\
    \hline\hline
   6  & 1 & 2 & $S_{LS}$ & 4 c + 2 $C_{LS}$  &  4 $S_{US}$  & 1 + 4c + 4$C_{US}$ \\
      & 2 & 1 & $O(\epsilon)$ & 2 + 4c  & $S_{US}$  & 4 + 4c + $C_{US}$ \\
   \hline\hline
      & Sum & 6 & $S_{LS}$ & 2 + 8c + 2 $C_{LS}$ &  5 $S_{US}$ & 5 + 8c + 5$C_{US}$ \\
   \hline\hline   
   8  & 1 & 3 & 6 $S_{LS}$ & 6c + 6$C_{LS}$ & 9 $S_{US}$ & 1 +6c + 9$C_{US}$ \\
      & 3 & 1 & $O(\epsilon)$ & 6 + 6c  & $S_{US}$  & 9 +6c + 9$C_{US}$ \\
      & 2 & 2 & 2 $S_{LS}$ & 2 + 8c +  2 $C_{LS}$  & 4 $S_{US}$ & 4 + 8c + 4 $C_{US}$ \\
   \hline\hline
      & Sum & 8 & 8 $S_{LS}$ & 8 + 20c + 8 $C_{LS}$ & 14 $S_{US}$ & 14 + 20c + 14 $C_{US}$ \\
   \hline
\end{tabular}
\end{center}
 \end{table}

Due to the reflection symmetry with respect to the $x$-axis, the charged hadron pairs from the $- x$ part of the overlapping
region have the same azimuthal correlations.
The total azimuthal correlations for the like-sign charged particles are
\begin{eqnarray}
 \langle \sin(\phi_{\alpha})\sin(\phi_{\beta})\rangle_{LS} &=& \frac{1}{Z} [P(4;\lambda)\,O(\epsilon)
+ P(6;\lambda)\,S_{LS} + 8 P(8, \lambda)\, S_{LS}], \label{ssls}\\
  \langle \cos(\phi_{\alpha})\cos(\phi_{\beta})\rangle_{LS} &=& \frac{1}{Z} [P(4,\lambda)\,2c 
+  P(6,\lambda)\,(2+8c+2C_{LS}) +  P(8, \lambda)\,(8+20c+8C_{LS})], \nonumber \label{ccls}\\
&&  
\end{eqnarray} 
where $Z$ is the normalization factor and $P(K,\lambda)$ is the Poisson distribution for $K$ particles
with a mean of $\lambda$. 

Likewise, the total azimuthal correlations for the unlike-sign charged particles are
\begin{eqnarray}
 \langle \sin(\phi_{\alpha})\sin(\phi_{\beta})\rangle_{US} &\!\!\!=\!\!\!& \frac{1}{Z} [P(4,\lambda)\,S_{US}
+ 5 P(6,\lambda)\,S_{US} + 14 P(8, \lambda)\,S_{US}], \label{ssus}\\
  \langle \cos(\phi_{\alpha})\cos(\phi_{\beta})\rangle_{US}  &\!\!\!=\!\!\!& \frac{1}{Z} [P(4,\lambda)\,(1+2c+C_{US})
+  P(6,\lambda)\,(5+8c+5C_{US}) +  P(8, \lambda)\,(14+20c+14C_{US})]. \nonumber \label{ccus}\\
&&
\end{eqnarray} 

There are few prominent features that we can readily observe from Eqs.(\ref{ssls})-(\ref{ccus}).
With the mean value being the average cluster size $\lambda=M_{cluster} = 3$, the Poisson probabilities are 
\mbox{$P(4;3) = 0.168$,} $P(6;3) = 0.0504, P(8;3) = 0.00810$ (We have ignored the 10-particle contribution due to its small
$P(10;3) = 0.00081$ which is an order of magnitude smaller than $P(8;3)$). The correlations are dominated by the case
with 4 particles. The noted features are:
\begin{itemize}
\item 
Since the 4-particle contribution to 
$ \langle \sin(\phi_{\alpha})\sin(\phi_{\beta})\rangle_{LS}$ is negligible (i.e. $O(\epsilon)$), it is expected to be small 
as compared to other correlations which is consistent with Eq.~(\ref{LSS}). 

\item 
Since $P(4;3)$ is the leading term and the fact that $c <0$, $\langle \cos(\phi_{\alpha})\cos(\phi_{\beta})\rangle_{LS}$ 
in Eq.~(\ref{ccls}) turns out to be negative. This is consistent with experiments in Eq.~(\ref{LCC}). 

\item We see from Eqs.~(\ref{ccus}) that if $c \sim -0.5 $ which is
close to the uncorrelated estimate of $- 2/\pi$, then the out-out correlations are largely canceled
out by the out-in cross correlations for each of the 4-, 6-, and 8-particle cases. Furthermore, we notice
that the coefficients of  $S_{US}$ and $C_{US}$ are identical in each case with different particle numbers.
Thus, if $S_{US} \sim C_{US}$ which is true for the flat distribution
of $f_{US}$ in Eq.~(\ref{S_US}), then the $sine$ and $cosine$ correlations of the unlike-sign pairs are
about equal, i.e. $\langle \sin(\phi_{\alpha})\sin(\phi_{\beta})\rangle_{US} \simeq 
\langle \cos(\phi_{\alpha})\cos(\phi_{\beta})\rangle_{US}$ which is what is found experimentally in Eq.~(\ref{USC}).
\end{itemize}
Thus, the present model with simplistic scenarios and assumptions seem to be able
to capture the generic features of the azimuthal correlations in Eqs.~(\ref{LSS}) - (\ref{USC}). 

To be more specific, let's take  
$c = -0.55$ and $C_{LS} = S_{US} = C_{US} = 0.8$ and $\lambda = M_{cluster} =3$,
we find 
\begin{equation}
\langle \cos(\phi_{\alpha})\cos(\phi_{\beta})\rangle_{LS} :
\langle \sin(\phi_{\alpha})\sin(\phi_{\beta})\rangle_{US}:
\langle \cos(\phi_{\alpha})\cos(\phi_{\beta})\rangle_{US} = -0.46:1:1.09
\end{equation}
These ratios are in quantitative agreement with those of experiments in the centrality range of 35\% - 65\% as 
illustrated in Fig. 2 in Ref.~\cite{bkl10}. 
The fact that $ \langle \sin(\phi_{\alpha})\sin(\phi_{\beta})\rangle_{US}$ and 
$\langle \cos(\phi_{\alpha})\cos(\phi_{\beta})\rangle_{US}$ are near identical is understood as
due to the cancellation between the out-out and out-in correlations for the unlike-sign charged particles. 
The fact that $\langle \sin(\phi_{\alpha})\sin(\phi_{\beta})\rangle_{US}$ is negative is due to the fact that
it is dominated by the out-in cross contribution (i.e. terms associated with $c$) which is negative. 

As for the smallness of $\langle \sin(\phi_{\alpha})\sin(\phi_{\beta})\rangle_{LS}$, we see that the ratio 
between Eqs.~(\ref{ssls}) and (\ref{ssus}) depends only on $\frac{S_{LS}}{S_{US}}$
\begin{equation}
\frac{\langle \sin(\phi_{\alpha})\sin(\phi_{\beta})\rangle_{LS}}
{\langle \sin(\phi_{\alpha})\sin(\phi_{\beta})\rangle_{US}} = 0.216\, (\frac{S_{LS}}{S_{US}}).
\end{equation}
If $\frac{S_{LS}}{S_{US}}$ is about 0.5, it would roughly explain the experimental finding that
$\frac{\langle \sin(\phi_{\alpha})\sin(\phi_{\beta})\rangle_{LS}}
{\langle \sin(\phi_{\alpha})\sin(\phi_{\beta})\rangle_{US}}$ is about an order of magnitude
smaller than $\langle \sin(\phi_{\alpha})\sin(\phi_{\beta})\rangle_{US}$ and
$\langle \cos(\phi_{\alpha})\cos(\phi_{\beta})\rangle_{US}$ at 60\% centrality. 

\section{Conclusion}   \label{conclusion}

    To conclude, we find the cluster model description with the assumption that the charge-dependent azimuthal correlations originate
mainly from the clusters on the surface of the overlapping region of the colliding nuclei and that the outward going charged hadron pairs
from these surface clusters are parallel to the reaction plane can describe the azimuthal
correlations of both the like-sign charged pairs and unlike-sign charged pairs in a qualitative manner. 
In particular, we find that the hard to understand near equality of the $sine$ and $cosine$ correlations of the unlike-sign pairs 
i.e. $\langle \sin(\phi_{\alpha})\sin(\phi_{\beta})\rangle_{US} \simeq \langle \cos(\phi_{\alpha})\cos(\phi_{\beta})\rangle_{US}$ is
due to the cancellation between the out-out pairs and the out-in cross correlated pairs. 
In addition, the centrality dependence of the correlations is naturally explained in terms of the cluster model~\cite{PHO-NN} and
the surface cluster dominance is suggested from the suppression of the cluster size of the near-side pairs as compared to the
away-side pairs in experiment~\cite{PHO-NN}. 
We also point out that there is a strong transient electric fields induced by the rapidly varying magnetic field in the
initial phase of the collision. Folding in the timeline of parton production, we find that most partons produced on the
surface during the semi-hard scattering after $t =0$ with mid rapidities coincide with the rapid decrease of the magnetic field and
will experience a momentum transfer as large as several hundred MeV from the induced electric field along the reaction plane. 
This could have some relevance to our assumption of the outward going charged hadron pairs.  
To verify this simplified picture presented here and explain the data more quantitatively, one needs
to incorporate the large transient magnetic field and its collateral electric field from the Faraday law in
the very early stage of parton production (0.1 fm/c) in dynamical models such as 
Hadron String Dynamics (HSD)~\cite{ec96}, HIJING~\cite{wg91}, AMPT~\cite{lkl05}, NeXus~\cite{dlo02},
momentum kick model~\cite{cywong11}, and 3D Relativistic Hydrodynamics~\cite{hag06}.
Only when the background contribution from the parity-non-violating QCD + electromagnetic effects are understood to enough 
precision will one be able to address the existence of 
the chiral magnetic effect in these azimuthal correlations.

This work is partially support by U.S. DOE Grants No. DE-FG05-84ER40154. The author wishes to thank H. Huang, 
F. Wang, \mbox{C.Y. Wong}, and N. Xu for useful discussions and particularly C.Y. Wong for reading the manuscript. 
He also thanks Mingyang Sun for drawing the figures for this manuscript.

\end{document}